\documentclass[aps,twocolumn,longbibliography]{revtex4-2}
\usepackage{color}
\usepackage{graphicx}
\usepackage{bm,bbm,amsmath,amssymb,amsthm,amsfonts}
\usepackage{enumerate}
\usepackage{latexsym}
\usepackage{tikz}  

\DeclareMathAlphabet\mathbfcal{OMS}{cmsy}{b}{n}
\usepackage{graphicx}
\usepackage{dcolumn}
\usepackage{array}
\usepackage{xcolor}
\newcolumntype{P}[1]{>{\centering\arraybackslash}p{#1}}
\usepackage{float}

\newcommand{\bs}{\boldsymbol}

\newcommand{\floor}[1]{\lfloor #1 \rfloor}
\newcommand{\ceil}[1]{\lceil #1 \rceil}

\usepackage[colorlinks=true,urlcolor=blue,citecolor=blue,allcolors=blue]{hyperref}

\begin{document}
	
	\title{Many-body entanglement and topology from uncertainties and measurement-induced modes}
	\author{Kim P\"oyh\"onen$^{1,2}$, Ali G. Moghaddam $^{1,2,3}$, Teemu Ojanen$^{1,2}$ }
	\affiliation{$^{1}$Computational Physics Laboratory, Physics Unit, Faculty of Engineering and
		Natural Sciences, Tampere University, P.O. Box 692, FI-33014 Tampere, Finland}
	\affiliation{$^{2}$Helsinki Institute of Physics P.O. Box 64, FI-00014, Finland}
	\affiliation{$^{3}$Department of Physics, Institute for Advanced Studies in Basic Science (IASBS), Zanjan 45137-66731, Iran}
	
	\begin{abstract}
		We present universal characteristics of quantum entanglement and topology through virtual entanglement modes that fluctuate into existence in subsystem measurements. For generic interacting systems and extensive conserved quantities, these modes give rise to a statistical uncertainty which corresponds to entanglement entropies. Consequently, the measurement-induced modes provide directly observable route to entanglement and its scaling laws. Moreover, in topological systems, the measurement-induced edge modes give rise to quantized and non-analytic uncertainties, providing easily accessible signatures of topology. Our work provides a much-needed direct method to probe the performance of emerging quantum simulators to realize entangled and topological states.
	\end{abstract}
	\maketitle
	
	\section{Introduction}
	Entanglement is a signature property of quantum systems and has been a source of active debate since the early days of quantum theory \cite{EPR1935,Schrodinger1935}. Remarkably, in the last decade, entanglement has become a central unifying concept throughout physics, from condensed matter and quantum information to quantum gravity and black holes \cite{Horodecki,vedral2008rmp,plenio2014introduction,calabrese2004entanglement,preskill2007,Srednicki1993,ryu2006,hartman2013time,Tatsuma2018holography}. In condensed matter, entanglement has been recognized as central classifying property of distinct quantum phases of matter and many-body dynamics, and phase transitions \cite{laflorencie2016quantum,Kitaev2006,Fidkowski2011,calabrese2007entanglement,eisert2015quantum,osterloh2002scaling,Kitaev2003}. Entanglement measures, such as entanglement entropy and the entanglement spectrum, have been very successful in analysing different aspects of topological order and exotic phenomena such as many-body localization \cite{wen2006,Haldane2008,Nussinov2009,Pollmann2010,Thomale2010,jiang2012identifying,Schoutens2007,Bardarson2012,Gogolin2011,Serbyn2013,Nayak2013,Abanin2019rmp,nandkishore2015many}. In general, the entanglement entropy of low-lying states of a gapped Hamiltonian follows the area law scaling while systems with Fermi surfaces obey a logarithmic volume law scaling \cite{hastings2007area,Eisert2010,Plenio2005,Zanardi2005,Wolf2006,Swingle2010,Eisert2010}. The entanglement scaling laws form the basis of powerful practical methods to simulate strongly correlated many-body systems \cite{Schollwock2005dmrg,Orus2014}.
	
	While entanglement has been recognized as a key to understanding formal aspects of many-body systems, its role in experiments has remained modest. Proposals to measure entanglement entropy and entanglement spectrum typically involve delicate system-specific requirements and setups \cite{Greiner2015measuring,Greiner2016quantum,Hafezi2016,dalmonte2018quantum}. For example, transport measurements are generally unsuitable for the emerging Noisy Intermediate-Scale Quantum (NISQ) computers and quantum simulator systems \cite{preskill2018NISQ,kandala2017hardware,bernien2017probing,zhang2017observation,Nori2014simulation,Bloch2017quantum}. Also, NISQ devices  present a new compelling reasons to experimentally probe entanglement. The necessary prerequisite to perform successful quantum simulation  is the ability to prepare and preserve many-body entangled states. Thus, for quantitative assessment of the performance of quantum simulators, it is imperative to design methods that allow feasible direct probing of many-body entanglement. 
	
	\begin{figure}
		\includegraphics[width=0.99\columnwidth]{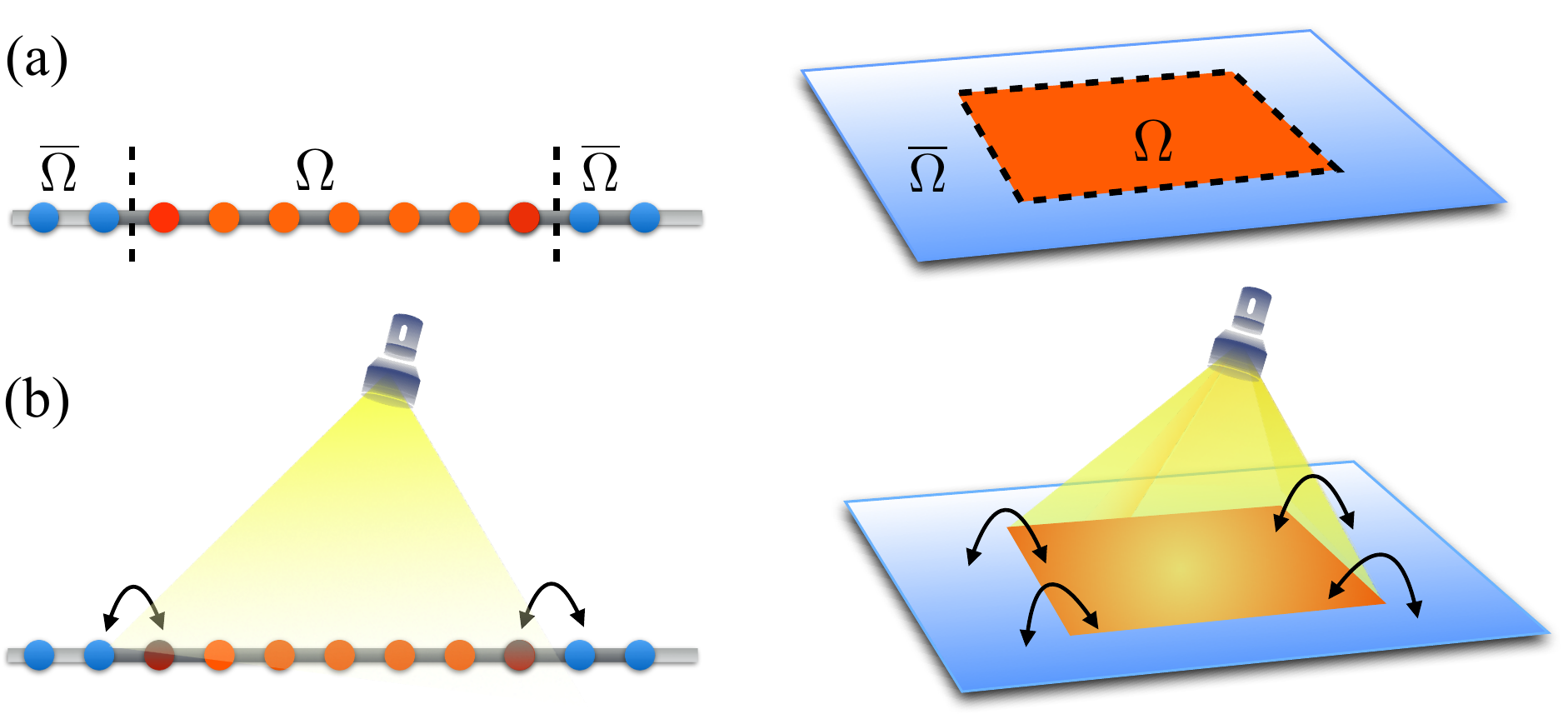}
		\caption{(a): Fictitious partitioning of a many-body system into subsystems $\Omega$ and $\overline\Omega$ is characterized by an entanglement spectrum with virtual entanglement edge (or bulk) modes. (b): Subjecting  subsystem $\Omega$ to an actual measurement of observable $\mathcal{A}$, the virtual entanglement modes become physical fluctuations. These contribute to the uncertainty $\delta^2 \mathcal{A}$ which, for conserved $\mathcal{A}$, provides a directly observable measure of entanglement. }
		\label{fig:schematic}
	\end{figure}

	In this work, we establish a protocol to extract many-body entanglement and topology by measurement-induced entanglement edge and bulk modes as illustrated in Fig.~\ref{fig:schematic}. This protocol is based on measurements of a collective quantity $\mathcal{A}$, such as total spin or particle number, of the subsystem $\Omega$. We show that for generic interacting systems, the statistical variance $\delta^2 \mathcal{A}$ of measurements of a conserved extensive quantity is a faithful measure of bipartite entanglement. The reason for this is that the uncertainty $\delta^2 \mathcal{A}$ solely originates from the virtual entanglement modes which fluctuate into existence due to the subsystem measurements. {We present a counting argument, supported by numerical calculations, which indicates} that $\delta^2 \mathcal{A}$ in general scales identically with entanglement entropies, thus providing a direct route to entanglement scaling laws. We illustrate that for strongly-correlated topological models, the measured uncertainty $\delta^2 \mathcal{A}$ exhibit quantized values as well as non-analytic behaviour which allow for direct extraction of many-body topology already from small subsystem measurements. Our work offers a directly observable route to measure entanglement and topology through measurement-induced entanglement modes.
	\par

	\section{General connection of uncertainty, entanglement and topology}
	
	Here we consider generic many-body systems in a pure state $|\Psi\rangle$ and introduce general relations between entanglement, topology and uncertainties in subsystem measurements. The starting point is the many-body entanglement spectrum $\{\lambda_i\} ,\{|\lambda_i\rangle\}$ which determines the reduced density matrix of the subsystem $\Omega$ as $\rho_{\,\Omega}=\mathrm{Tr}\,_{\overline\Omega}\,|\Psi\rangle\langle \Psi|= \sum_i\lambda_i|\lambda_i\rangle\langle\lambda_i|$ and subsystem operator $\mathcal{A}$ which represents a collective observable on $\Omega$. Here $\mathcal{A}$ could be arbitrary, but the physically most interesting results follow for observables that correspond to a sum of local operators, such as the total spin  for spin systems and the number of particles for fermionic systems. Subjecting subsystem $\Omega$ to measurement of $\mathcal{A}$, the expectation value and the variance/uncertainty are given by $\langle \mathcal{A}\rangle=\sum_i\lambda_i\langle \mathcal{A}\rangle_i $ and $\delta^2 \mathcal{A}=\langle \mathcal{A}^2\rangle-\langle \mathcal{A}\rangle^2$, where $\langle \mathcal{A}\rangle_i=\langle\lambda_i|\mathcal{A} |\lambda_i\rangle$. In contrast to the expectation value, which is simply a sum of expectation values of local operators, the uncertainty contains information on non-local correlations and entanglement. As shown in the Appendix \ref{App:I}, the uncertainty for a general operator satisfies inequality
	\begin{align}\label{eq:var1}
		\delta^2\,\mathcal{A}\geq \frac{1}{2}\sum_{i,j} \lambda_i\lambda_j\left(\langle\mathcal{A}\rangle_i-\langle\mathcal{A}\rangle_j\right)^2,
	\end{align}
	where the equality is satisfied by conserved quantities defined by $\big[\mathcal{A},\rho_{\,\Omega}\big]=0$. From now on, we focus on conserved quantities. Since $\mathcal{A}$ and $\rho_{\,\Omega}$ can be simultaneously diagonalized, the entanglement modes $|\lambda_i\rangle$ can be labelled by eigenvalues $\mathcal{A}_i$. The uncertainty of conserved observables is minimal, arising purely due to fluctuations between the entanglement modes $\lambda_i$ in different measurements, and vanish for unentangled states satisfying $\lambda_1=1$ and $\lambda_i=0$ for $i>1$.  Denoting the set of distinct eigenvalues as $\{\mathcal{A}_i\}$, we can express Eq.~\eqref{eq:var1} as summation over sectors of entanglement spectrum corresponding to different $\mathcal{A}_i$ as 
	\begin{align}\label{eq:var2}
		\delta^2\,\mathcal{A}=  \sum_{\mathcal{A}_i >\mathcal{A}_j}
		\lambda_{\mathcal{A}_i}\lambda_{\mathcal{A}_j}\left(\mathcal{A}_i-\mathcal{A}_j\right)^2,
	\end{align}
	where $\lambda_{\mathcal{A}_i}=\sum_{\langle \mathcal{A}\rangle_i=\mathcal{A}_i}\lambda_i$ is the probability to observe outcome $\mathcal{A}_i$. The set
	$\{\lambda_{\mathcal{A}_i}\}$ can be regarded as a coarse-grained entanglement spectrum filtered by the outcomes of $\mathcal{A}$. The relation \eqref{eq:var2} allows us to make a connection between the measured uncertainty and the bipartite entanglement. 
	As we will see in the following, such a connection can be based on generic counting arguments along with the general properties of the entanglement spectrum. 
	
	\subsection{Identical scaling of uncertainties and entanglement entropies}
	\label{subsec:general-scaling}
	 Here, we provide a physical counting argument that the uncertainty \eqref{eq:var2} for observables such as total spin $S_{z,\Omega}=\sum_{i\in \Omega} S^z_i$ or particle number $N_{\Omega}=\sum_{i\in \Omega}\hat{n}_i$, which consists of sums of local observables, and for which the number of distinct outcomes scale as the subsystem degrees of freedom, exhibits the same scaling as von Neumann and R\'enyi entanglement entropies. Moreover, the utility of entanglement entropies as entanglement measures typically stems from their scaling as ${N}_{a}$, which denotes the number of degrees of freedom in the subsystem that efficiently participate in the entanglement. For example, when a lattice spin system exhibits an area-law  (volume-law) entanglement, ${N}_{a}$ corresponds to a number of spins on the surface layer of finite thickness (spins in the whole volume). When ${N}_{a}$ spins in a subsystem are effectively entangled with its environment, the entropies scale as ${\cal S}_2\sim {\cal S}_{vN}\sim {N}_{a}$. This follows from the fact that the $2^{{N}_{a}}$ different microstates (distinct states in the entanglement spectrum) each have a nonzero probability of the order of $\lambda_i\sim (1/2)^{{N}_{a}}$, thus 
		\begin{equation}
			{\cal S}_2=-\ln \sum_i\lambda_i^2\sim -\sum_i\lambda_i\ln \lambda_i = {\cal S}_{vN} \sim {N}_{a}.
			\label{eq:Renyi-scaling}    
		\end{equation}

		Next, we consider conserved extensive quantities, such as the total subsystem spin $\mathcal{A}=\sum_{i\in \Omega} S^z_i$ or particle number $\mathcal{A}=\sum_{i\in \Omega}\hat{n}_i$, and outline the fundamental reason why their uncertainty given by
		Eq. \eqref{eq:var2} obeys the same scaling laws as the entanglement entropies. We first note that for extensive observables, the number of distinct outcomes $n_i$ (for example, the total spin or the number of fermions in the subsystem) scales as the subsystem volume or 
		its total number of sites $N_{\Omega}$. For a conserved quantity, fluctuations arise solely from the variation of spin configurations between different states in the entanglement spectrum. Depending on whether the state of the system displays an area-law or volume-law entanglement, the effective number of fluctuating spins $N_a$ correspond to either spins in the surface layer or in the whole system.  Importantly, when we consider extensive observables that are sum of local observables, the macrostate probabilities  $\lambda_{{\cal A}_i}$ are strongly peaked regardless of the details of the microstate distribution $\lambda_i\sim 1/2^{{N}_{a}}$. For instance, in spin $1/2$ or spinless fermion systems, the number of configurations for total spin $S_z=n_i-{N}_{a}/2$ or particle number $n_i$ is given by the binomial coefficient $\binom{{N}_{a}}{n_i}$. Consequently, the distribution of ${\cal A}$  is approximated by the binomial distribution $\lambda_{n_i}\sim \binom{{N}_{a}}{n_i} \frac{1}{2^{{N}_{a}}}$ whose average and variance are proportional to $N_{a}$. Thus, \emph{uncertainties for a conserved extensive sum variable $\mathcal{A}$ scale as ${N}_{a}$, in agreement with entanglement entropies}. Similar reasoning applies also for observables that can take more than two (but finite) number of values per lattice site by considering multinomial coefficients and distributions. 
		
		The above explanation outlines that the observable $\mathcal{A}$ will display the same scaling for uncertainties as entropies, as long as it is conserved, extensive and a sum of independent variables.
		As noted above, the usefulness of entropies as entanglement measures mostly derives from the fact that they scale as the effective number of entangled degrees of freedom between the subsystems. Therefore, remarkably, for an extensive conserved sum of local operators, $\delta^2\,\mathcal{A}$ \emph{can be conceptually regarded as an entanglement measure on equal footing with entropies}. As a crucial advantage over entropies, a bipartite entanglement and the entanglement scaling laws can be directly measured through uncertainties, as demonstrated below for various paradigmatic spin and fermion models. This property enables experimental observation of entanglement  in wide range of emerging quantum simulator systems. 
	
	\subsection{Signature of topology in uncertainties}
	Topological order of a many-body state implies a special structure of the entanglement spectrum, which also becomes directly observable through uncertainties. The entanglement spectrum is conveniently discussed in terms of the entanglement Hamiltonian $H_E=-\ln{\rho_{\Omega}}$, which shares the topological properties of the studied system. A gapped topological state generally obeys the entanglement area law, with a dominant part of the entanglement arising from low-lying states of $H_E$ which represent topological edge modes. As illustrated in Fig.~\ref{fig:schematic}(b), subsystem measurement induces edge-mode fluctuations that determine $\delta^2\,\mathcal{A}$ of a conserved quantity. This effect is particularly striking for 1d systems, where the entanglement spectrum exhibits topological ground state degeneracy between states that differ only by their edge mode configurations. As shown in Appendix \ref{App:I}, the contribution of the topological edge modes to uncertainties can be estimated using the exact bound 
	\begin{align}\label{eq:var_subspace}
		\delta^2\,\mathcal{A}\geq
		\lambda_\Sigma^2  \: \delta^2_{\Sigma}\,\mathcal{A},
	\end{align}
	where $\lambda_{\Sigma}=\sum_{i\in\Sigma}\lambda_i$ is the probability to find the system in a low-lying subspace $\Sigma$ of the entanglement Hamiltonian $H_E$ and $\delta^2_{\Sigma}\,\mathcal{A} $ is calculated using the density matrix $\rho_\Sigma = \sum_{i\in \Sigma} (\lambda_i/\lambda_\Sigma)\,| \lambda_i  \rangle \langle \lambda_i | $ restricted into $\Sigma$. The formula is valid for general systems and does not require that the low-lying manifold exhibit exact degeneracies. However, in 1d systems away from critical points, the (nearly) degenerate ground state manifold dominates, so that with exponential accuracy we get $\lambda_{\Sigma}\to 1$ and $\delta^2\,\mathcal{A}= \delta^2_{\Sigma}\,\mathcal{A}$ which is constant in the topological phase. As illustrated by examples below, in 1d systems this gives rise to quantized uncertainty plateaus determined by the fluctuating edge modes. Furthermore, in generic topological systems, topological phase transitions can be observed through striking non-analytic features of $\delta^2\,\mathcal{A}$ near critical points due to the abrupt appearance of the edge modes in the entanglement spectrum.

	\begin{figure}
		\includegraphics[width=0.99\columnwidth]{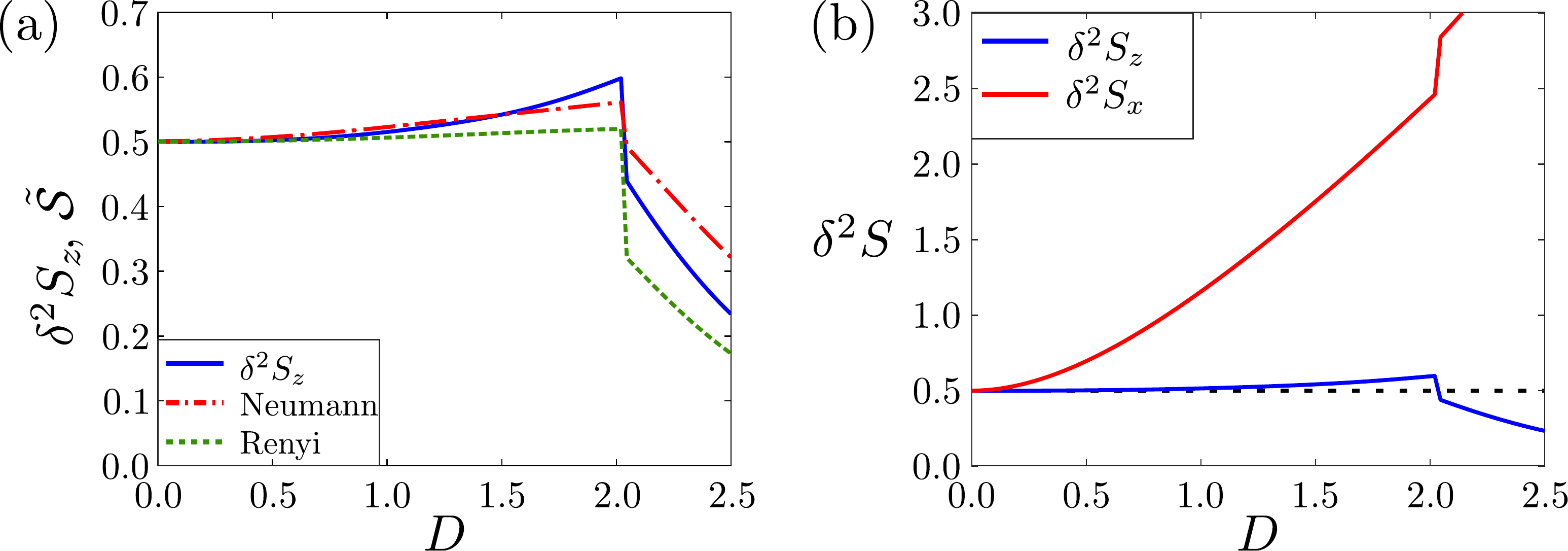}
		\caption{Uncertainties and entropies in the AKLT chain as a function of the local spin anisotropy for the total system length $L=15$ (periodic boundary conditions) and subsystem $N=7$ \cite{lengthfootnote}. (a): Comparison between $\delta^2 S_z$ and entanglement entropies, with the latter scaled to $\tilde{\mathcal S} =  \mathcal S/(4\ln 2)$. (b): Uncertainties of subsystem spin components $\delta^2 S_z$ and $\delta^2 S_x$. The conserved component exhibits a quantized plateau $\delta^2 S_z=0.5$ due to the measurement-induced spin-$\frac{1}{2}$ edge modes. 
		}
		\label{fig:spin_one}
	\end{figure} 
	
	\section{Strongly-correlated systems}
	Here we illustrate the general connection between uncertainties entanglement, topology and measurement-induced states with paradigmatic models of correlated systems. These examples explicitly demonstrate the intimate relationship between uncertainties and entanglement entropy. 
	
	\subsection{Spin 1 chain}
	The first model is the Affleck-Kennedy-Lieb-Tasaki (AKLT) spin-1 chain $\mathcal{H}=\mathcal{\hat H}_0+\mathcal{\hat H}_1$, with
	\begin{equation} \label{eq:spinone}
		\mathcal{\hat H}_0 =\sum_{\nu,j = 1}^{L} J_\nu\hat S_j^\nu\hat S_{j+1}^\nu+\alpha\sum_{j = 1}^{L} (\hat {\bs{S}}_j \cdot \hat {\bs{S}}_{j+1})^2
	\end{equation} 
	and $\mathcal{\hat H}_1=D\sum_{j = 1}^L(\hat{S_j^z})^2$, where $D>0$ is a local spin anisotropy. For isotropic $J_x=J_y=J_z$ couplings and $\alpha=\frac{1}{3}$ the ground state is an exactly solvable valence bond state. The AKLT chain \eqref{eq:spinone} supports the celebrated topological Haldane phase down to the isotropic Heisenberg point $\alpha=0$. The topological phase is present at small $D$ but is destroyed for large $D$ in favour of the local singlet state. For isotropic exchange couplings, the $z$ component of the total spin is conserved. This implies that for any subsystem $\Omega$, the total spin $S_{z,\Omega}=\sum_{i\in \Omega} S^z_i$ is conserved $\big[S_{z,\Omega},\rho_{\,\Omega}\big]=0$. Thus, the bipartite entanglement can be obtained by measuring the uncertainty $\delta^2 S_z$ for a subsystem. In Fig.~\ref{fig:spin_one}(a) we compare $\delta^2 S_z$ to von Neumann entropy ${\mathcal S}_{vN}=-\sum_i\lambda_i\ln{\lambda_i}$ and the 2nd R\'enyi entropy ${\mathcal S}_{2}=-\ln{\sum_i\lambda_i^2}$. Apart from normalization, $\delta^2 S_z$ and the entropies provide essentially the same information. Deep in the topological phase, they are constant as required by the entanglement area law. When approaching the critical point, where the system becomes gapless, all three quantities exhibit a cusp that signifies a crossover from the area law to a critical Luttinger-type behaviour. At the critical point, the uncertainty and the cusp exhibit discontinuity, after which the entanglement is strongly suppressed due to entering the local singlet-dominated phase.  Figure~\ref{fig:spin_one}(b) illustrates the comparison between different subsystem spin components $\delta^2 S_z$, which is conserved, and $\delta^2 S_x$ which is not conserved. Here we see that it absolutely \emph{crucial to consider a conserved quantity} as a measure of entanglement. Only $\delta^2 S_z$ is sensitive to entanglement and exhibits the quantized plateau deep in the topological phase. The quantized value $\delta^2 S_z=0.5$ can be directly understood in terms of measurement-induced end states. The topological end modes in the Haldane phase correspond to spin 1/2 excitations at each end of the subsystem. These two states combine to a triplet state $S=1$ with $S_z=-1,0,1$ (each with probability $\lambda_i=1/4$) and singlet state $S=0,S_z=0$ (with probability $1/4$). Thus, the quantized uncertainty $\delta^2 S_z=0.5$ has a straightforward interpretation that, when imposing a measurement, each edge spin configuration fluctuates into existence with probability $1/4$.
	
	{
		The above example provides a particular clear illustration of the generic mechanism how the observed uncertainty in subsystem quantities is directly caused by measurement-induced entanglement modes. The entanglement spectrum and the associated edge modes arise due to a virtual partitioning of the system. Before a measurement, the boundaries between the two subsystems have no physical significance since the system is homogeneous. Thus, the edge modes in the entanglement spectrum are purely fictitious and have no independent physical standing. Now, performing a subsystem measurement, say an approximately projective measurement of a conserved quantity, provides the subsystem boundaries a clear physical importance which did not exist before a measurement. After a measurement, the subsystem is observed in one of the eigenstates of the reduced density matrix with the corresponding probability. When the system is deep in a topological phase, as seen in the above example, the subsystem states after a measurement differ due to their edge spin configurations. Therefore, 
		the virtual edge modes of the entanglement spectrum acquire a concrete physical existence in the subsystem measurements. For conserved quantities, these measurement-induced modes are the source of the observed uncertainty, which also provides a faithful measure of bipartite entanglement.  
		
	}
	
	\subsection{Gapped spin 1/2 chain}

	\begin{figure}
		\includegraphics[width=0.99\columnwidth]{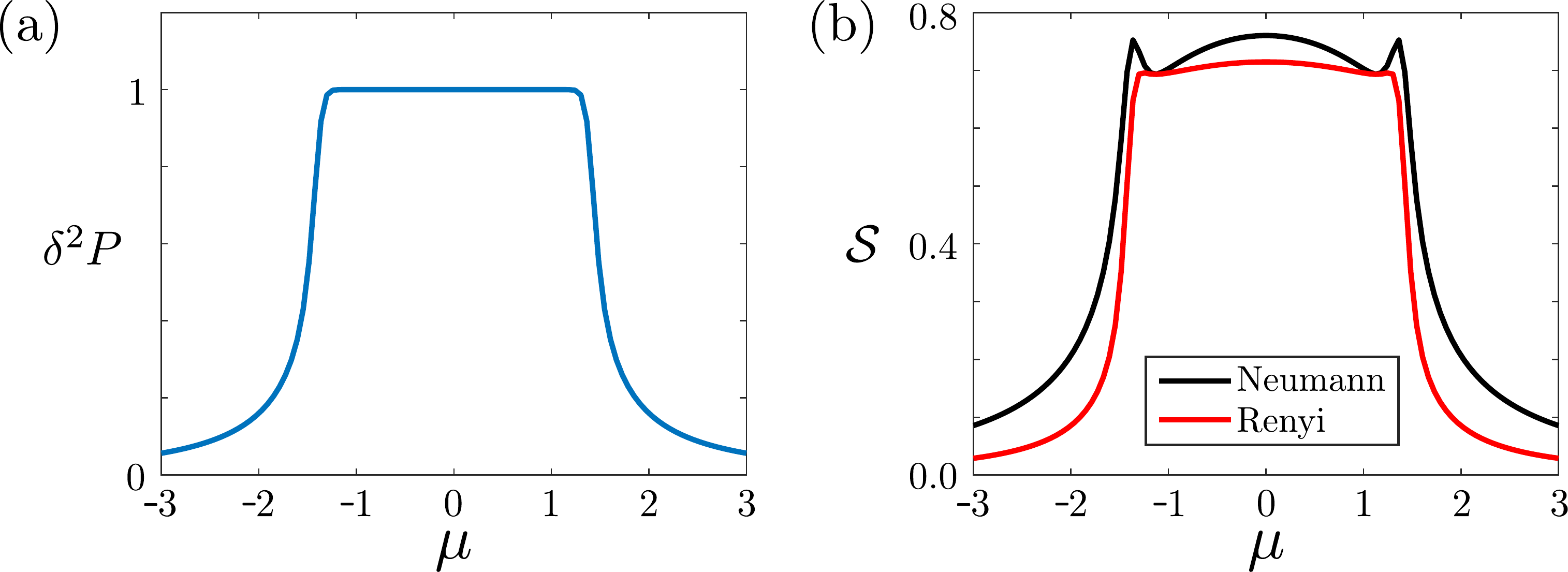}
		\caption{Uncertainty of the subsystem parity (a) and the entropies (b) for the interacting Kitaev chain as a function of the chemical potential for $\Delta/t=0.9$, $V/\Delta=0.5$ with $V > 0$, the system length $L=20$ and subsystem $N=10$.
		}
		\label{fig:spin_half}
	\end{figure} 
	For the second example, we consider a spin-$\frac{1}{2}$ XYZ chain in magnetic field $\mathcal{\hat{H}}=\mathcal{\hat H}_0+\mathcal{\hat H}_2$, where $\mathcal{\hat H}_0$ is the same as in Eq.~\eqref{eq:spinone} with $\alpha=0$ and the spin 1 operators replaced by spin 1/2 operators, and $\mathcal{\hat H}_2=h\sum_j\hat{S_j^z}$. With the Jordan-Wigner mapping, the spin model is transformed to the interacting Kitaev chain of spinless fermions
	\begin{equation}
		\label{eq:kitaev}
		\begin{split}
			\mathcal{\hat{H}} =& -\frac{1}{2}\sum_{j = 1}^{L-1}\left(t\,\hat {c}_{j}^\dagger\hat{c}_{j+1} + \Delta\, \hat {c}_{j}^\dagger\hat{c}_{j+1}^\dagger + \text{H.c.}\right)\\ 
			&+V\sum_{j = 1}^{L-1} \left(\hat {n}_{j}-\frac{1}{2}\right)\left(\hat {n}_{j+1}-\frac{1}{2}\right)-\mu\sum_{j = 1}^{L} \hat {c}_{j}^\dagger\hat{c}_{j},
		\end{split}
	\end{equation}
	where parameters of the two Hamiltonians are related as $t=-(J_x+J_y)/2$, $\Delta=(J_y-J_x)/2$, $\mu=-h$, and $V=J_z$. The spin system has a discrete spin rotation symmetry which translated to conservation of electron parity in the fermion model. The Kitaev chain is the paradigmatic model of topological superconductivity. In the topological small $|\mu|$ regime, a system with open boundaries harbours two Majorana end modes which can accommodate a single non-local fermion excitation and give rise to a doubly degenerate ground state. While the particle number conservation is violated by the superconducting pairing terms, the parity of particles is conserved. Defining the parity operator $P$ so that it takes value $1$ (-1) when the subsystem particle number is even (odd), the entanglement and topology can be readily extracted from the uncertainty $\delta^2P$. In the spin language, the parity is defined through the parity of the number of up spins in the subsystem. The quantized value $\delta^2P=1$, again, simply reflects the measurement-induced edge modes: the non-local fermion excitation has 0.5 probability to be populated after each measurement. Figure~\ref{fig:spin_half}(a) illustrates how the quantized plateau in the parity uncertainty reveals the topological phase diagram already for small systems. This clearly reflects the behaviour of entanglement entropies seen in Fig.~\ref{fig:spin_half}(b), for which the plateau also becomes clearly resolved. 
	
	\subsection{Critical spin 1/2 chain}
	\begin{figure}
		\includegraphics[width=0.8\columnwidth]{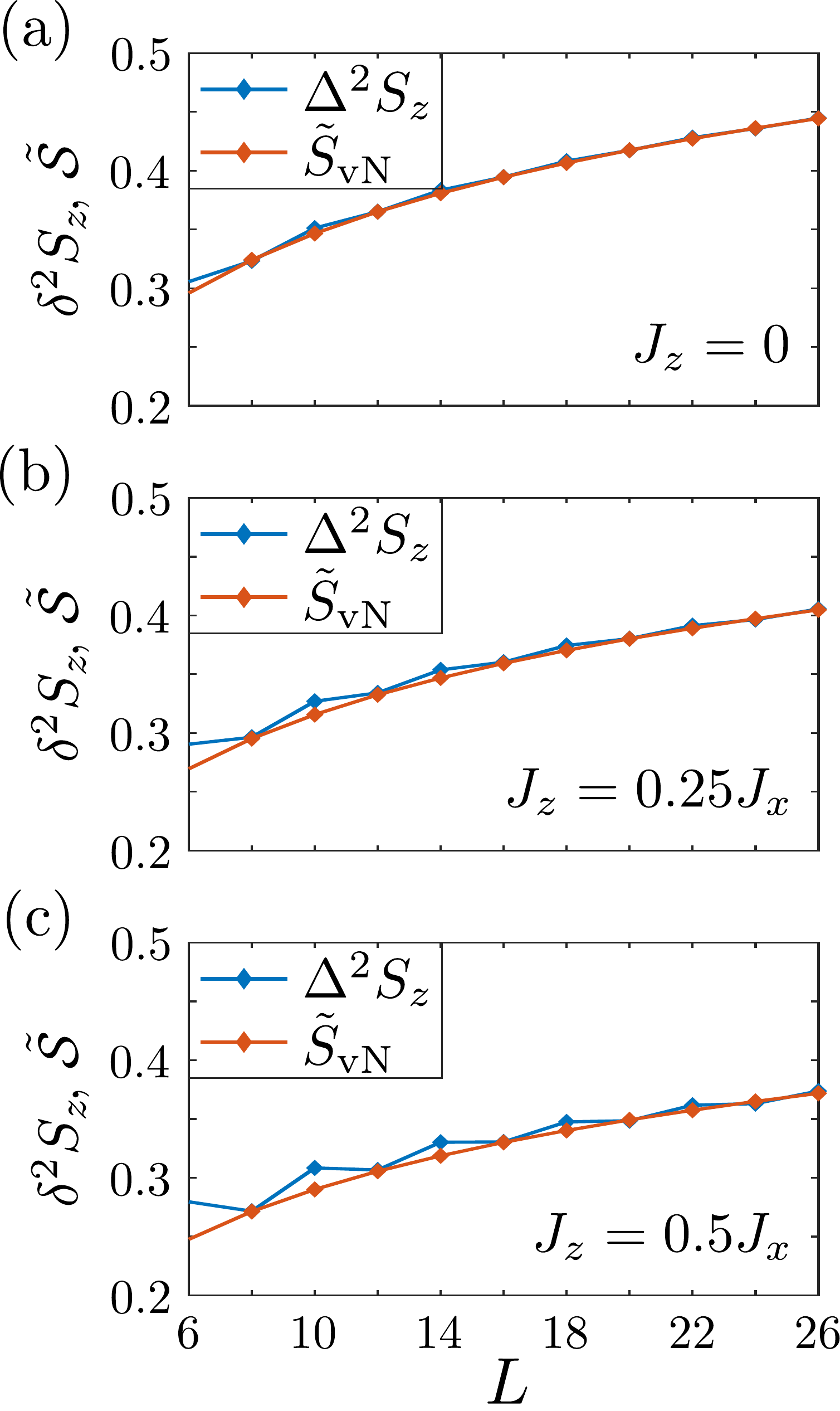}
		\caption{
			Identical size scaling of the spin fluctuations and the von Neumann entropy in a critical XXZ spin chain, shown as a function of system length $L$ for subsystems of length $L_s=L/2$. Different panels correspond to different values of $J_z$. By increasing $J_z$ the fluctuations are overall slightly suppressed but the length dependence remains intact. 
			The von Neumann entropy
			has been scaled 
			with numerical factors (a) 3.22, (b) 3.54, and (c) 3.86 in different panels.
		}
		\label{fig:spinhalf-scaling}
	\end{figure} 
	
		The gapped spin chains studied above demonstrated the match between entanglement and fluctuations in the area-law regime where they do not depend on the subsystem length. Here we demonstrate that the identical scaling behavior of entropy and spin fluctuations $\Delta S_z^2$ also hold in the critical regime of the spin-$1/2$ XXZ chain ($J_x=J_y$).
		In the absence of magnetic field and when $|J_z/J_x|\leq 1$, the XXZ spin chain enters a gapless phase where the entanglement entropy scales logarithmically with the system length \cite{CalabreseCardy2004,LeHur2010}. 
		As illustrated in Fig.~\ref{fig:spinhalf-scaling}, the variance of spin fluctuations $\Delta S_z^2$ and the von Neumann entanglement entropy reveal almost identical scaling behavior with length. The case $J_z=0$ in (a) maps to free fermions while panels (b) and (c) correspond to strongly interacting fermions. The overall size scaling of entropy and uncertainties is in excellent agreement. We see that increasing $J_z$ gives rise to finite-size oscillations around the mean depending on the parity of the subsystem size. However, these finite-size effects are seen to vanish for larger sizes and we get a perfect match between the uncertainties and scaled von Neumann entropy $\tilde{\cal S}_{vN}$. Our results have been obtained using exact diagonalization, so the known log-scaling cannot be extracted meaningfully from the available system sizes. Nevertheless, the match between the uncertainties and the entanglement entropy provides strong evidence for the identical scaling of entropies and uncertainties in critical chains. To summarize this section, the results obtained in interacting area law and critical chains are in excellent agreement with our general counting argument presented in Sec.~\ref{subsec:general-scaling}.

	\section{Free fermion systems}Here we illustrate the general theory in free Fermi systems. In systems with particle number conservation, the information about the subsystem is encoded in the correlation matrix ${\bm C}$ with elements ${C}_{ij}^{\alpha\alpha'}=\langle\hat{c}^\dagger_{i\alpha}\hat{c}_{j\alpha'}\rangle$ where $\hat{c}^\dagger_{i\alpha},\hat{c}_{i\alpha}$ are creation and annihilation operators of a particle at site $i$ with orbital (internal degrees such as spin) index $\alpha$, respectively \cite{Peschel_2003}. 
	Considering a general single-particle operator ${\mathcal{A}}=\sum_{i,j,\alpha,\alpha'} A_{ij}^{\alpha,\alpha'} \hat{c}^\dagger_{i\alpha}\hat{c}_{j\alpha'} $ of the subsystem, we can show that the uncertainty reads (see Appendix \ref{App:II} for the derivation),
	\begin{equation}
		\label{eq:variance-free-ferions}
		\delta^2 {\mathcal{A}}={\rm Tr}\left[{\bm A}^T {\bm C} {\bm A}^T ({\mathbbm 1}-{\bm C}) \right],    
	\end{equation}
	which reduces to the known result $\delta^2 {N}_{\Omega}={\rm Tr}({\bm C}-{\bm C}^2)$ for the particle number fluctuations \cite{poyhonen2021}.

	\begin{figure}
		\includegraphics[width=0.9\columnwidth]{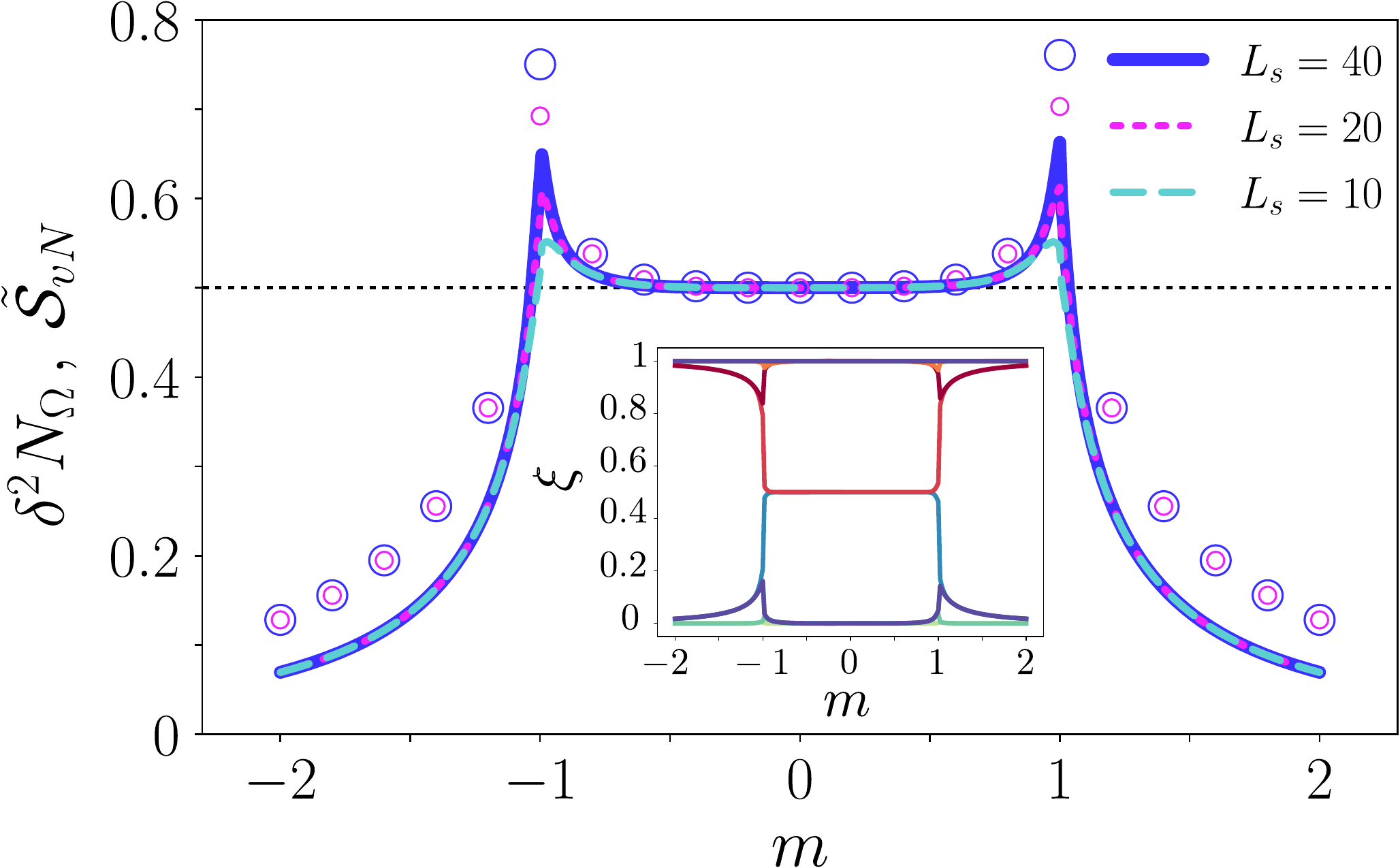}
		\caption{Topology-enforced uncertainties in 1D system. 
			When the system is in the topological phase ($|m|<1$), the particle fluctuations exhibit a quantized plateau stemming from measurement-induced end modes. Open circles indicate the entanglement entropy for $L_s=20$ and 40 for comparison. Inset shows the eigenvalues of the correlation matrix.}
		\label{fig:topo-variance}
	\end{figure}

\begin{figure}[t!]
		\includegraphics[width=0.99\columnwidth]{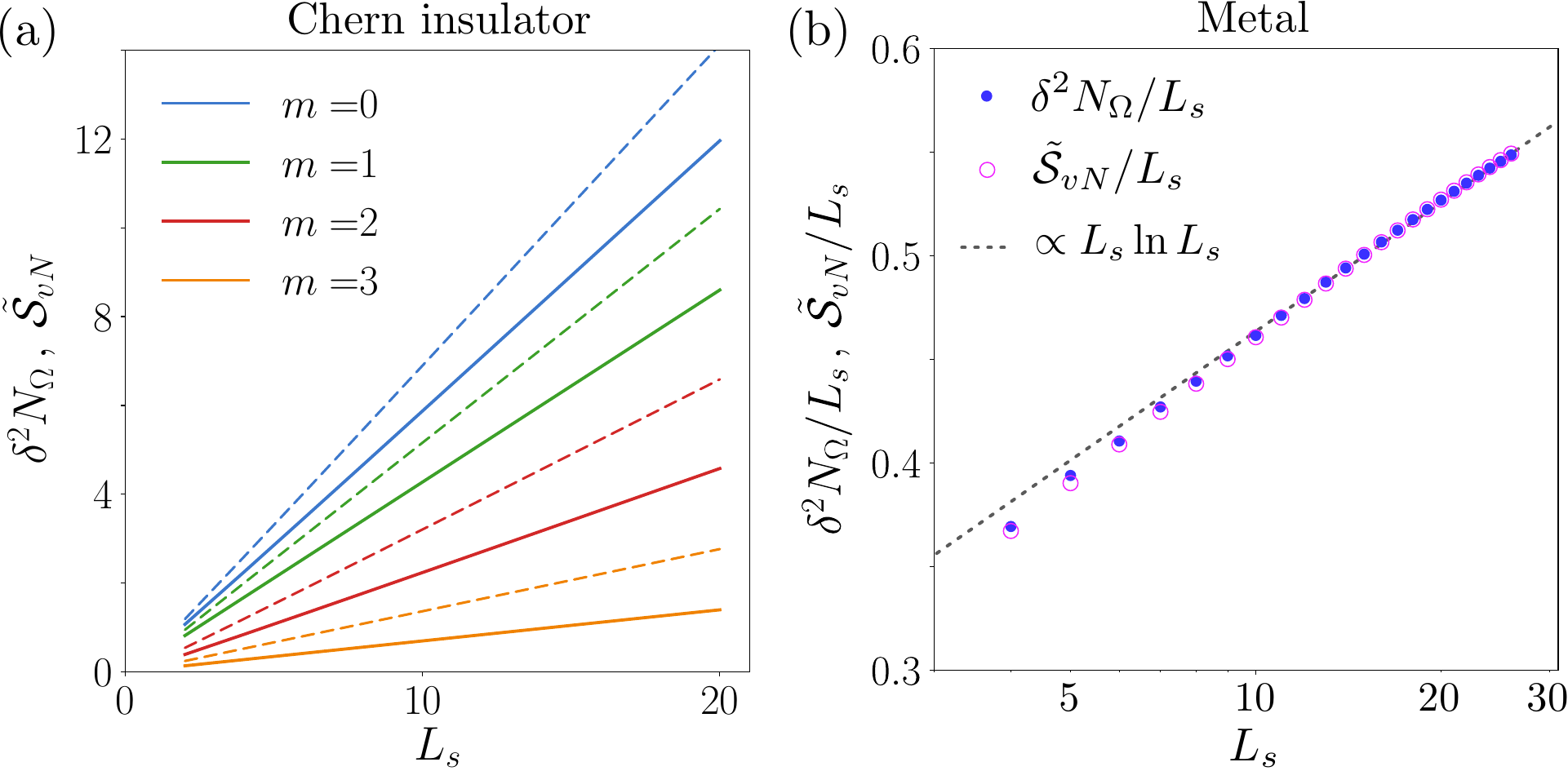}
		\caption{Identical scaling laws for the particle fluctuations and the von Neumann entanglement entropy. (a) Area law scaling for 2D Chern insulators, with solid and dashed lines representing particle fluctuations and the entropy, respectively. (b) Log-volume law for 2D metal. The von Neumann entropy has been scaled with with a factor $1.165$.}
		\label{fig:scaling}
	\end{figure}

	First, we demonstrate the correspondence of $\delta^2 {N}_{\Omega}$ and entanglement entropies for a 1D topological insulator with Hamiltonian $\hat{\cal H}_{1D}=(m-\cos k)\sigma_z+\sin k\sigma_x$ as shown in Fig. \ref{fig:topo-variance}. Particularly, in the topological phase ($|m|<1$), 
	$\delta^2 {N}_{\Omega}$ and the scaled entropy $\tilde{\cal S}_{vN}={\cal S}_{vN}/(4\ln2)$ coalesce and form a plateau at a quantized value $1/2$. Also, in the trivial phase ($|m|>1$) we see a non-universal (yet length-independent) behavior for both $\delta^2 {N}_{\Omega}$ and $\tilde{\cal S}_{vN}$ which diminish when $|m|\gg1$. At the transition point $|m|=1$ though, we see a length-dependent cusp, indicating a well-known gapless Luttinger liquid behavior $\log L_s$-dependence at the critical point \cite{calabrese2004entanglement}. Next, we examine similarity of scaling laws for other prototypical models, namely the 2D Chern insulator ${\cal H}_{\rm QWZ}=(m-\cos k_x-\cos k_y)\sigma_z+\sin k_x \sigma_x+\sin k_y \sigma_y$ and a 2D single-band metal with dispersion $\varepsilon_k=-2t(\cos k_x+\cos k_y)$. Figure~\ref{fig:scaling}(a) shows that both $\delta^2 N$ and $\tilde{\cal S}_{vN}$ 
	reveal area-law scaling behavior in the 2D Chern model, which persists even at the Dirac-type gap closing points $m=0,\pm2$. However, for the metal with a finite Fermi surface, Fig.~\ref{fig:scaling}(b) shows a logarithmic volume-law scaling for both quantities in agreement with the general form ${\cal S}_{vN} \sim L_s^{d-1}\ln L_s$ for a $d$-dimensional metal \cite{Klich2006Widom,Wolf2006,Swingle2010,LeHur2012PRL,LeHur2012}

	Finally, we generalize the results for charge fluctuations and show that fluctuations of arbitrary conserved quantities provide a direct access to topology and entanglement scaling laws in a more general sense. Considering conserved quantities ${\mathcal{A}}$, for which $[{\mathcal{A}},\rho_{\Omega}]=0$, we find that their uncertainties are bounded by the number fluctuation as $ {A}^2_{\rm min} \delta^2 N_{\Omega} \leq   \delta^2 {\mathcal{A}}\leq  {A}^2_{\rm max} \delta^2 N_{\Omega} $, where ${A}_{\rm max,min}$ denote the largest and smallest magnitude eigenvalue of the single-particle matrix ${\bm A}^T$. For typical observables of interest, ${A}_{\rm max,min}$ does not depend on the system size, which means that ${\mathcal{A}}$ obeys the same size scaling as $\delta^2 N_{\Omega}$, which implies $\delta^2 {\mathcal{A}}\propto{\cal S}_{vN}$ for a conserved ${\cal A}$. In contrast, for nonconserved observables such as spin $S_z$ in 2D Chern model or the total subsystem energy, in general obey a volume-law scaling $L^d$.
	Detailed derivation of these features can be seen in Appendix \ref{App:II}.

	\section{Discussion and outlook} 
	We established that a statistical uncertainty in conserved extensive observables, arising from measurement-induced 
	entanglement modes, allows a direct experimental access to many-body entanglement and topology. In particular, the uncertainties provide a promising direct probe of entanglement properties in emerging NISQ devices and quantum simulators. Since the necessary prerequisite to carry out successful quantum simulations include the ability to prepare and preserve desired entangled many-body states, it is crucial to benchmark the entanglement performance of the devices. Our proposed protocol provides a direct method to carry out this task through simple measurements. As our results  clearly indicate that, already in the modest-sized programmable devices with $\lesssim 20$ functional qubits that are now becoming accessible for wide audiences \cite{heim2020quantum}, one could directly and accurately probe entanglement area laws and topology through uncertainties. 
	
	To conclude, we highlight the conceptual and practical differences between our work and the previous works on the connection between entanglement entropy and fluctuations.  Rooted in the free fermion transport problems and full-counting statistics \cite{Klich2006,LeHur2012}, these works establish that the R\'enyi entanglement entropy for free fermions can be exactly expressed in terms of the electronic cumulants \cite{LeHur2012,LeHur2012PRL,LeHur2010} and discuss to what extent a similar result could apply to interacting systems by considering two-point correlations in systems with conserved quantities. As concluded in Ref.~\cite{LeHur2012}, no general connection between the entropy and correlations was obtained beyond the free fermion systems. Instead of correlation functions or specific Hamiltonians, our starting point is based on the general structure of the entanglement spectrum and its labelling in terms of conserved quantum numbers. This approach leads us to the counting argument which reveals the connection between the directly measurable uncertainties and
	entanglement entropy, manifesting in their identical scaling behaviour. 
	
	Also, the physical and experimental implications of our work are very different than previous related works \cite{LeHur2012,LeHur2012PRL,LeHur2010}. We propose a scheme to experimentally access entanglement and topology by measuring a non-local observable such as total spin or particle number of the subsystem. In this scheme, in contrast to measuring separately a set of two-point functions, a single measurement addresses the subsystem as whole and introduces physical boundaries. Our protocol, as highlighted with the spin systems, gives rise to physical measurement-induced boundary and entanglement modes which can even lead to quantization of the observed uncertainties.
	We believe that, due to the impressive progress in various quantum simulator systems, 
	measuring conserved subsystem observables is quite feasible even in current setups. 
	The conserved quantities are, for instance, the number of particles inside the subsystem for fermionic models or projections of the total spin in spin systems. In particular, we note the existing experiments where
	the particle statistics inside a subsystem has been already measured for small subsystems \cite{Greiner2016quantum}.

	In upcoming work, we will generalize the present work to quantifying multipartite entanglement, far-from-equilibrium entanglement dynamics and aspects of the entanglement spectrum from directly measurable uncertainties.

	\acknowledgements
	The authors acknowledge the Academy of Finland project 331094 for support.

	\appendix

	\section{General variance relations}
	\label{App:I}
	As discussed in the main text, the statistical variation (uncertainty) in the outcomes of subsystem measurements have general properties summarized in Eqs.~(1-3) in the main text.
	We begin proving inequality (1), by writing
	\begin{align}
		\delta^2 \mathcal{A}&=\langle \mathcal{A}^2 \rangle - \langle \mathcal{A} \rangle^2 
		=\sum_{i} \lambda_i\langle\mathcal{A}^2\rangle_i-\Big(\sum_{i} \lambda_i\langle\mathcal{A}\rangle_i\Big)^2 \nonumber \\
		&=\sum_{i,j} \lambda_i\lambda_j\langle\mathcal{A}^2\rangle_i-\sum_{i,j} \lambda_i\lambda_j\langle\mathcal{A}\rangle_i\langle\mathcal{A}\rangle_j,
		\label{eq:var_new2}
	\end{align}
	where $\sum_j\lambda_j=1$ has been used to obtain the last expression. By symmetrizing the first term, we obtain the inequlity
	\begin{align}
		\delta^2\,\mathcal{A}&=
		\frac{1}{2}  \sum_{i,j} \lambda_i\lambda_j  \left(    \langle\mathcal{A}^2\rangle_i+\langle\mathcal{A}^2\rangle_j -2 \langle \mathcal{A}_i\rangle \langle\mathcal{A}\rangle_j   \right)\nonumber\\
		&\geq \frac{1}{2} \sum_{i,j} \lambda_i\lambda_j\left(\langle\mathcal{A}\rangle_i-\langle\mathcal{A}\rangle_j\right)^2,
		\label{eq:var_new3}
	\end{align}
	which follows from the relation $\langle\mathcal{A}^2\rangle_i\geq \langle\mathcal{A}\rangle_i^2$ for arbitrary state $|\lambda_i\rangle$. 
	Note that for a conserved quantity $\mathcal{A}$, which is simultaneously diagonalized in the $|\lambda_i\rangle$ basis, we have $\langle\mathcal{A}^2\rangle_i = \langle\mathcal{A}\rangle_i^2$. Thus, in the case of conserved $\mathcal{A}$, the general relation \eqref{eq:var_new3} actually becomes equality.
	
	Next we derive an rigorous uncertainty bound by considering only states in an arbitrary subspace, for example, the low-lying subspace of the entanglement Hamiltonian. Since each term of the summation on both sides of relation \eqref{eq:var_new3} is individually non-negative, the full expression is never smaller than the corresponding one with $i,j$ restricted to an arbitrary subspace $\Sigma$ of the full entanglement spectrum. Hence we can write
	\begin{align}
		\delta^2\,\mathcal{A}&\geq  \sum_{i,j\in\Sigma}  \lambda_i  \lambda_j \Big( \langle\mathcal{A}^2\rangle_i - \langle\mathcal{A}\rangle_i\langle\mathcal{A}\rangle_j     \Big)
		\nonumber \\
		&= 
		\lambda_\Sigma^2 \Big[
		\sum_{i\in\Sigma}  \frac{\lambda_i}{\lambda_\Sigma}\:\langle\mathcal{A}^2\rangle_i-\Big(\sum_{i\in\Sigma} \frac{\lambda_i}{\lambda_\Sigma}\: \langle\mathcal{A}\rangle_i\Big)^2\Big],
		\label{eq:var_general_1}
	\end{align}
	where we have defined the accumulative probability $\lambda_\Sigma=\sum_{i\in\Sigma} \lambda_i$ for the subsystem being in subspace $\Sigma$. This proves Eq.~(3) in the main text since the expression inside the brackets in the right-hand side above is nothing but the uncertainty $\delta^2_{\Sigma}\,\mathcal{A}$ obtained in the truncated subspace.

	\section{Uncertainty formulas for free Fermi systems}
	\label{App:II}
	Pioneered by Peschel \cite{Peschel_2003,Peschel_2009}, the entanglement spectrum for free fermions in a Gaussian state can be completely obtained from the correlation matrix elements ${C}_{ij}^{\alpha\alpha'}=\langle\hat{c}^\dagger_{i\alpha}\hat{c}_{j\alpha'}\rangle$. Subsequently and by applying wick's theorem, expectation value of any observable can be described in terms of the correlation matrix ${\bm C}$.
	So for a general single-particle operator ${\mathcal{A}}=\sum_{i,j,\alpha,\alpha'} A_{ij}^{\alpha,\alpha'} \hat{c}^\dagger_{i\alpha}\hat{c}_{j\alpha'} $ we have
	\begin{widetext}
		\begin{align}
			\delta^2 {\mathcal{A}} 
			&=\sum_{ijkl}\sum_{\alpha\alpha'\beta\beta'}
			A_{ij}^{\alpha\alpha'} A_{kl}^{\beta\beta'}
			\Big(
			\langle  \hat{c}^\dagger_{i\alpha}\hat{c}_{j\alpha'} \hat{c}^\dagger_{k\beta}\hat{c}_{l \beta'} \rangle
			-
			\langle  \hat{c}^\dagger_{i\alpha}\hat{c}_{j\alpha'}\rangle \langle \hat{c}^\dagger_{k\beta}\hat{c}_{l \beta'} \rangle
			\Big)
			=
			\sum_{ijkl}\sum_{\alpha\alpha'\beta\beta'}
			A_{ij}^{\alpha\alpha'} A_{kl}^{\beta\beta'}
			\langle  \hat{c}^\dagger_{i\alpha} \hat{c}_{l \beta'} \rangle \langle
			\hat{c}_{j\alpha'} \hat{c}^\dagger_{k\beta} \rangle \nonumber \\
			&=\sum_{ijkl}\sum_{\alpha\alpha'\beta\beta'}
			A_{ij}^{\alpha\alpha'} A_{kl}^{\beta\beta'}
			C_{il}^{\alpha\beta'}  
			\Big(\delta_{jk}\delta_{\alpha'\beta}-C_{kj}^{\beta\alpha'}\Big) 
			= 
			{\rm Tr}\left[{\bm A}^T {\bm C} {\bm A}^T ({\mathbbm 1}-{\bm C}) \right],\label{eq:general-uncertanity}
		\end{align}
	\end{widetext}
	which can be recast as $\delta^2 {\mathcal{A}} = \sum_{\xi\xi'}\xi(1 - \xi')\: |\langle\xi|{\bm A}^T | \xi'\rangle|^2$ using the eigenstates $| \xi \rangle$ of the ${\bm C}$ matrix. For conserved observables, we get a simpler result $\delta^2 {\mathcal{A}} = \sum_{\xi}\xi(1 - \xi)\: |\langle\xi|{\bm A}^T | \xi\rangle|^2$  
	and particularly $\delta^2 N_\Omega= \sum_{\xi}\xi(1 - \xi) $ for the total fermionic number of the subsystem. The positivity of each terms in these expressions immediately force the general bounds $ {A}^2_{\rm min} \delta^2 N_{\Omega} \leq   \delta^2 {\mathcal{A}}\leq  {A}^2_{\rm max} \delta^2 N_{\Omega} $, where ${A}_{\rm max,min}$ denote the largest and smallest magnitude eigenvalue of  ${\bm A}^T$. This result in combination with the exact bound $\delta^2 N_\Omega\leq{\mathcal{S}}_{vN}/(4\ln2)$ dictated by the entanglement entropy \cite{Klich2006}, suggests the similar scaling behavior of any conserved observable and entanglement entropies. Particularly, as illustrated explicitly for prototypical models in the main text, we see area law and
	logarithmic volume law for conserved quantities in gapped and metallic phases, respectively.
	
	Next, we show that non-conserved observables are, in general, limited by a volume-law scaling as the upper bound, namely
	\begin{align}
		\delta^2 {\mathcal{A}} \leq {\rm Tr}\left[ {\bm C} \big({\bm A}^T\big)^2 \right] \leq A_{\rm max}^2  \ceil{\langle N_{\Omega} \rangle}.\label{eq:nonconserved-volume-law}
	\end{align}
	The first inequality can be deduced from Eq. \eqref{eq:general-uncertanity} by throwing the negative term $\big[({\bm A}^T {\bm C}\big)^2  \big]$ away. For the second part of the relation, we first note 
	${\rm Tr}\big[ {\bm C} \big({\bm A}^T\big)^2 \big] =   \sum_j  \langle A_j | {\bm C} | A_j\rangle\, A_j^2$
	using the eigenstates $| A_j\rangle$ of ${\bm A}^T$. Since each term in both sequences $A_j^2$ and $p_j\equiv\langle A_j | {\bm C} | A_j\rangle$ are non-negative, re-ordering the two sequences such that both become ascending (we can assume $|A_j|\leq|A_{j+1}|$ and re-order $p_j$'s such that $\tilde{p}_j\leq \tilde{p}_{j+1}$) the sum of their pairwise multiplications becomes larger, thus ${\rm Tr}\big[ {\bm C} \big({\bm A}^T\big)^2 \big]\leq \sum_j \tilde{p}_j \: A_j^2$. Now, we notice that $\sum_{j =1}^J \tilde{p}_j ={\rm Tr}\,{\bm C} = \langle N_\Omega \rangle $, and consequently
	\begin{align}
		\sum_{j =1}^{J- \ceil{ \langle N_\Omega \rangle }} \tilde{p}_j  \leq \sum_{j =J- \floor{ \langle N_\Omega \rangle }}^{J} \big(1-\tilde{p}_j\big),
	\end{align}
	with $J$ denoting the size of the subsystem Hilbert space (equivalently maximum number of fermions allowed). But exploiting the sortedness the two sequences $A_j^2$ and $\tilde{p}_j$ once again, we further obtain 
	\begin{align}
		\sum_{j =1}^{J-  \ceil{ \langle N_\Omega \rangle }} \tilde{p}_j\:A_j^2 \: \leq \sum_{j =J- \floor{ \langle N_\Omega \rangle }}^{J} (1-\tilde{p}_j) \:A_j^2,
	\end{align}
	and then 
	\begin{align}
		\sum_{j =1}^{J} \tilde{p}_j\:A_j^2& \leq \sum_{j =J- \floor{ \langle N_\Omega \rangle }}^{J} A_j^2
		\leq  A_{\rm max}^2 \: \ceil{ \langle N_\Omega \rangle },  
	\end{align}
	which completes the proof of the volume-law inequality \eqref{eq:nonconserved-volume-law}.

	\bibliography{subentanglement}

\begin{thebibliography}{60}
\bibitem [{len()}]{lengthfootnote}%
  \BibitemOpen
  \href@noop {} {\ }\bibinfo {note} {In an antiferromagnetic chain, the
  finite-size results exhibit notable dependence on the parity of $L$ which is
  odd in this example. The parity dependence, which vanishes in the
  thermodynamic limit, can be minimized by applying twisted boundary conditions
  for the total system.}\BibitemShut {Stop}%
\end{thebibliography}
	
\end{document}